\documentclass[leqno,12pt,draft]{amsart}
\usepackage{bm}
\usepackage{amssymb}
\numberwithin{equation}{section}
\theoremstyle{plain} % italic
\newtheorem{theorem}{\indent\bf Theorem}[section] 
\newtheorem{lemma}[theorem]{\indent\bf Lemma}
\newtheorem{corollary}[theorem]{\indent\bf Corollary}

\theoremstyle{definition} % 

\newtheorem{remark}[theorem]{\indent\bf Remark}

\begin{document}
\begin{flushright}
\today
\end{flushright}
\title[Exponential polynomials and Quantum computing]
{On exponential polynomials and quantum computing} %

\author[Y. Ohno]{Yasuo Ohno$^*$} % 

\author[Y. Sasaki]{Yoshitaka Sasaki$^\dagger$}

\author[C. Yamazaki]{Chika Yamazaki$^\dagger$}

%%%%%%%%%%%%%%%%%%% ‹r' %%%%%%%%%%%%%%%%%%%%%%%%%%%%%%
%\subjclass[2000]{ % 2000MSC
%Primary 00; Secondary 00.}

\thanks{%Thanks
$^*$The first author is supported in part by JSPS Grant-in-Aid No.~20540033.}
%
%Partly supported by the Grant-in-Aid for Scientific Research (*), 
%Japan Society for the Promotion of Science. }
% ‰ÈŒ¤"ï•â•@Šî"ÕŒ¤‹†(A)'̏ꍇ}
\thanks{$^\dagger$The second and third authors are supported in part by 
``Open Research Center" Project 
for Private Universities: matching fund subsidy from MEXT}

%%%%%%%%%%%% affiliation %%%%%%%%%%%%%

\address{
Department of Mathematics \endgraf
Kinki University \endgraf
Higashi-Osaka, Osaka 577-8502 \endgraf
Japan
}
\email{ohno@math.kindai.ac.jp}

\address{ 
Interdisciplinary Graduate School of Science and Engineering \endgraf
Kinki University \endgraf
Higashi-Osaka, Osaka 577-8502 \endgraf
Japan}
\email{sasaki@alice.math.kindai.ac.jp}

\address{ 
Department of Mathematics \endgraf
Kinki University \endgraf
Higashi-Osaka, Osaka 577-8502 \endgraf
Japan
}
\email{chika@math.kindai.ac.jp}

%%%%%%%%%%%%%%%%%%%%%%%%%%%%%%%%%%%%%%%%%%%%%%%%%%%%%%%

\maketitle

\begin{abstract}
We calculate the zeros of an exponential polynomial of three variables 
by a classical algorithm and quantum algorithms which are 
based on the method of van Dam and Shparlinski, they 
treated the case of two variables, 
and compare with the time complexity of those cases. 
Further we compare the case of van Dam and Shparlinski with 
our case by considering the ratio (classical$/$quantum) of the time complexity. 
Then we can observe the ratio decreases. 
\end{abstract}

%%%%%%%%%%%%%%%%%%%%%%%%%%%%%%%%%%%%%%%%%%%%%%%%
\section{Introduction}
%%%%%%%%%%%%%%%%%%%%%%%%%%%%%%%%%%%%%%%%%%%%%%%%%%%%%%%%
For a prime number $p$, we put $q = p^{\nu}$, where $\nu$ is 
a certain positive integer. 
Then we denote the finite field by $\mathbb{F}_q$ which has $q-1$ elements. 
Namely, $\mathbb{F}_q$ forms an additive group and 
$\mathbb{F}_q^{\times} := \mathbb{F}_q \backslash \{ 0 \}$ forms a 
multiplicative group, where $0$ is the zero element in $\mathbb{F}_q$. 
Any element of $\alpha \in \mathbb{F}_q^{\times}$ have a periodicity, that is 
there exits a smallest natural number $s$ such that $\alpha^s =1$. We call 
such $s$ the ``multiplicative order" of $\alpha$. It is known that 
the multiplicative order is a divisor of $\# \mathbb{F}_q^{\times} = q-1$.

To calculate the number of the zeros of a polynomial 
\[ F (x_1, \dots, x_{m}) = \sum_{(n_1, \dots, n_{m}) \in \mathbb{N}_0^m} 
a_{n_1, \dots, n_{m}} x_1^{n_1} \cdots x_{m}^{n_{m}} \]
is a very important problem in mathematics. Here, 
$\mathbb{N}_0 := \mathbb{N} \cup \{ 0 \}$ and 
$a_{n_1, \dots, n_{m}} \in \mathbb{F}_q$. 
In \cite{ds}, van Dam and Shparlinski treated 
the following exponential polynomial 
\begin{equation}
f (x, y) = a_1 g_1^{x} + a_2 g_2^{y} -b \label{eq2}
\end{equation}
and calculated the zeros of \eqref{eq2} by quantum algorithms. 
Further they compared the time complexity due to a classical algorithm with 
that due to a quantum algorithm. Then the ``cubic" speed-up was observed. 

In this article, we treat 
the following exponential polynomial 
\begin{equation}
f_b (x_1, x_2, x_3) := a_1 g_1^{x_1} + a_2 g_2^{x_2} + a_3 g_3^{x_3} -b \label{eq}
\end{equation}
and calculate the solutions of $f_b (x_1, x_2, x_3) = 0$ by 
using quantum algorithms which are natural generalizations of the method 
of van Dam and Shparlinski. 
Here, $a_i$, $g_j \in \mathbb{F}_q^{\times}$ ($i, j =1, 2, 3$) and 
$b \in \mathbb{F}_q$. Further we also compare the time complexity 
due to a classical algorithm with 
that due to a quantum algorithm. Then exponentially ``$5/2$ times" speed-up 
is observed. 

In the next section, we introduce some notation and give the considerable 
lemma which supports whether there exit the zeros of \eqref{eq}. 
In Section \ref{cl}, we evaluate the time complexity due to a classical algorithm. 
Further in Section \ref{qu}, we evaluate the time complexity due to 
a quantum algorithm. 

%%%%%%%%%%%%%%%%%%%%%%%%%%%%%%%%%%%%%%%%%%%%%%%%%%%%%%%%%%%%
\section{The number of solutions of equation}
%%%%%%%%%%%%%%%%%%%%%%%%%%%%%%%%%%%%%%%%%%%%%%%%%%%%%%%%%%%%%%
In this section, we give an important formula with respect to 
the density of solutions of 
\begin{equation}
f_b (x_1, x_2, x_3) := a_1 g_1^{x_1} + a_2 g_2^{x_2} + a_3 g_3^{x_3} -b = 0
\label{th1-eq}
\end{equation}
as Lemma \ref{lem1}, below. 
To state it, we introduce some notation. 

Let each $s_i$ be the multiplicative order of 
$g_i$ ($i=1, 2, 3$) in \eqref{th1-eq}. 
We put 
\begin{align*}
X_i &:= \{ 0, 1, \dots, s_i-1 \} \cong \mathbb{Z}/s_i \mathbb{Z}, \quad 
\text{($i=1, 2, 3$),} \\
X_{3} (r) &:= \{ 0, 1, \dots, r-1 \} \subseteq X_3 \quad 
\text{$(r = 1,2, \dots, s_3)$,} \\
\bm{X}^3 (r) &:= X_1 \times X_{2} \times X_{3} (r) 
\intertext{and}
\bm{X}^3 &:= \bm{X}^3 (s_3) = X_1 \times X_{2} \times X_{3}. 
\end{align*}
Then we define
\begin{align*}
S_{f_b} (r) &:= \{ (x_1, x_2, x_3) \in \bm{X}^3 (r) \ | \ 
f_b (x_1, x_2, x_3) = 0 \}, \\
N_{f_b} (r) &:= \# S_{f_b} (r) 
\end{align*}
for $r = 1, \dots, s_3$. 

By using above notation, we can state the following result:

\begin{lemma} \label{lem1}
Let $\delta$ be a parameter satisfying $\delta = o(q)$. 
For $r > \delta^2 q^3 (s_1 s_2)^{-2}$, 
we have 
\begin{equation}
N_{f_b} (r) = \frac{s_1 s_2 r}{q} 
+ O (\delta \sqrt{r q}), 
\end{equation}
except for at most $q/\delta^2$ exceptional $b$'s. 
Further $O$-constant can be taken $1$. 
\end{lemma}

Choosing $\delta = (\log q)^{1/2}$ in Lemma \ref{lem1}, we have 
\begin{corollary} \label{cor}
If $q^3 (s_1 s_2)^{-2} \log q < r \leq s_3$, 
then we see that $S_{f_b} (r) \neq \phi$ holds except for 
at most $q/\log q$ exceptional $b$'s. 
\end{corollary}

\begin{remark}
The above lemma and corollary make the point that 
the solutions of \eqref{th1-eq} exit only when 
\[ \frac{s_1 s_2}{q} \geq \Bigl( \frac{q}{s_3-2} \log q \Bigr)^{1/2} (> 1). \]
This inequality implies that the multiplicative orders $s_1$ and $s_2$ are 
somewhat large. 
\end{remark}

\begin{remark}
The exponent $1/2$ of $\delta = (\log q)^{1/2}$ is not necessary. 
In fact, $\delta = (\log q)^{\varepsilon}$ 
with any $\varepsilon > 0$ is sufficient. 
\end{remark}

\begin{proof}[\indent\sc Proof of Lemma \ref{lem1}]
Let $\psi$ be a non-trivial additive character over $\mathbb{F}_q$, in fact, any additive 
character over $\mathbb{F}_q$ can be given as a map $\mathbb{F}_q \to \mathbb{C}_1^*$, where 
$\mathbb{C}_1^* := \{ z \in \mathbb{C} | |z| = 1 \}$ (see \cite[Theorem 5.7]{ln}). 
To evaluate $N_{f_b} (\bm{v})$, we use the following formula which plays as a counting function:
\begin{equation}
\frac{1}{q} \sum_{\mu \in \mathbb{F}_q} \psi (u\mu) = 
\begin{cases}
1 & \text{if $u = 0$,} \\
0 & \text{otherwise.}
\end{cases} \label{or}
\end{equation}
Then we have
\begin{align}
N_{f_b} (r) &= \sum_{(x_1, x_2, x_3) \in \bm{X}^3 (r)}
\frac{1}{q} \sum_{\mu \in \mathbb{F}_q} 
\psi (\mu (f_b (x_1, x_2, x_3))) \label{nf} \\
&= \frac{s_1 s_2 r}{q} + \frac{1}{q} \sum_{\mu \in \mathbb{F}_q^*} 
\sum_{(x_1, x_2, x_3) \in \bm{X}^3 (r)} 
\psi (\mu (f_b (x_1, x_2, x_3))) \notag \\
&=: \frac{s_1 s_2 r}{q} + \Delta_b (r). \notag
\end{align}
If the contribution from the second term on the right-hand side of 
the above formula 
can be estimated by $o(s_1 s_2 r/q)$, the above formula tells us 
the existence of the solution of $f_b (x_1, x_2, x_3)$. 
To consider it, we evaluate the mean value of the second term on the right-hand 
side of \eqref{nf} 
with respect to $b$. Namely, we evaluate 
\[ E (r) := \sum_{b \in \mathbb{F}_q} 
\left| \Delta_b (r) \right|^2. \]
From \eqref{or} and some properties of the additive character over $\mathbb{F}_q$, 
we obtain
\begin{align*}
E (r) =& \frac{1}{q^2} \sum_{\mu, \mu' \in \mathbb{F}_q^{\times}} 
\left( \prod_{j=1}^2 \left( \sum_{x_j, x_j' \in X_j} 
\psi (a_j (\mu g_j^{x_j} - \mu' g_j^{x_j'})) \right) \right)
\sum_{x_3, x_3' \in X_3 (r)} 
\psi (a_3 (\mu g_3^{x_3} - \mu' g_3^{x_3'})) \\
& \quad \times \sum_{b \in \mathbb{F}_q} \psi (b(\mu'-\mu)) \\
=& \frac{1}{q} \sum_{\mu \in \mathbb{F}_q^{\times}} 
\left( \prod_{j=1}^2 \left( \sum_{x_j, x_j' \in X_j} 
\psi (a_j \mu (g_j^{x_j} - g_j^{x_j'})) \right) \right) 
\sum_{x_3, x_3' \in X_3 (r)} 
\psi (a_3 \mu (g_3^{x_3} - g_3^{x_3'})) \\
=& \frac{1}{q} \sum_{\mu \in \mathbb{F}_q^{\times}} 
\left( \prod_{j=1}^2 \Biggl| \sum_{x_j \in X_j} 
\psi (a_j \mu g_j^{x_j}) \Biggr|^2 \right)
\Biggl| \sum_{x_3 \in X_3 (r)} \psi (a_3 \mu g_3^{x_3}) \Biggr|^2. 
\end{align*}
It is known that 
\begin{align*}
\Biggl| \sum_{x_j \in X_j} \psi (a_j \mu g_j^{x_j}) \Biggr| 
&\leq \sqrt{q} \quad \text{for $j=1,2$ and any $\mu \in \mathbb{F}_q^{\times}$} 
\end{align*}
(see Theorem 8.78 in \cite{ln}). Hence we have
\begin{align*}
E (r) <& q
\sum_{\mu \in \mathbb{F}_q} 
\Biggl| \sum_{x_3 \in X_3 (r)} \psi (a_3 \mu f^{x_3}) \Biggr|^2 
= q^2 r. 
\end{align*}
Therefore, if we put $\delta = o(q)$, then we can see that 
there exit at most $q/\delta^2$ exceptional $b$'s such that 
\begin{equation}
\left| \frac{1}{q} \sum_{\mu \in \mathbb{F}_q^*} 
\sum_{(x_1, x_2, x_3) \in X_3 (r)} 
\psi (\mu (f_b (x_1, x_2, x_3))) \right| \geq \delta \sqrt{rq}. 
\end{equation}
Hence we obtain 
\begin{equation*}
N_{f_b} (r) = \frac{s_1 s_2 r}{q} 
+ O ( \delta \sqrt{qr} )
\end{equation*}
for other $b$'s. 
Now, the proof of Lemma \ref{lem1} is completed. 
\end{proof}

%%%%%%%%%%%%%%%%%%%%%%%%%%%%%%%%%%%%%%%%%%%%%%%%%%%
\section{Calculation of the deterministic time for a classical algorithm} \label{cl}
%%%%%%%%%%%%%%%%%%%%%%%%%%%%%%%%%%%%%%%%%%%%%%%%%%%%%%
We follow the method of van Dam and Shparlinski~\cite{ds}. Then we have 
\begin{theorem} \label{th1}
Except for at most $q/\log q$ exceptional $b$'s, 
we can either find a solution $(x_1, x_2, x_3) \in \bm{X}^3$ of the equation  
\eqref{th1-eq} or decide that it does not have a solution in deterministic time 
$q^{3/2} (\log q)^{O(1)}$ as a classical computer. 
\end{theorem}

\begin{proof}
Using a standard deterministic factorization algorithm, we factorize $q-1$ 
and find the orders $s_j$ ($j=1,2,3$) of $g_j$ in time $q^{1/2} (\log q)^{O(1)}$. 
We may assume without loss of generality that $s_1 \geq s_2 \geq s_3$. 
For calculated orders $s_1$ and $s_2$, we put 
\begin{equation}
r = \lceil q^3 (s_1 s_2)^{-2} \log q \rceil. \label{r}
\end{equation}
Then we see that the solution of \eqref{th1-eq} certainly exists 
when $r \leq s_3$. However, when $r > s_3$, 
we do not know whether such solutions exits. Therefore we have to 
consider those two cases. 

For each $(x_2, x_3) \in X_2 \times X_3 (r)$, 
we calculate the deterministic time of the discrete logarithm $x_1$ such that 
$g_1^{x_1} = a_1^{-1} ( b-a_2 g_2^{x_2} -a_3 g_3^{x_3})$. 
It is known that the deterministic time for this case is 
$s_1^{1/2} (\log q)^{O(1)}$ (see Section 5.3 in \cite{cp}). 
\begin{enumerate}
\item The case $r \leq s_3$. 
We have 
\[ (s_2 r) s_1^{1/2} (\log q)^{O(1)} \ll q^{3/2} (\log q)^{O(1)}, \]
since $s_1^{1/2} s_2 r < (s_1^2 s_2^2 r)^{1/2}$.
\item The case $r > s_3$. Similarly, we see that the deterministic time is 
\[ (s_2 s_3) s_1^{1/2} (\log q)^{O(1)} \ll q^{3/2} (\log q)^{O(1)}, \]
since $s_1^{1/2} s_2 s_3 < (s_1^2 s_2^2 s_3)^{1/2} < (s_1^2 s_2^2 r)^{1/2}$. 
\end{enumerate}
\end{proof}

%%%%%%%%%%%%%%%%%%%%%%%%%%%%%%%%%%%%%%%%%%%%%%%%%%%%%%%%%%%%%%%%%%%%%%%%
\section{Calculation of the time complexity for a quantum algorithm} \label{qu}
%%%%%%%%%%%%%%%%%%%%%%%%%%%%%%%%%%%%%%%%%%%%%%%%%%%%%%%%%%%%%%%%%%%%%%%%
In this section, we describe quantum algorithms which are based on the 
method of \cite{ds}. 

\begin{theorem} \label{th2}
Except for at most $q/\log q$ exceptional $b$'s, 
we can either find a solution $(x_1, x_2, x_3) \in \bm{X}^3$ of the equation 
\eqref{th1-eq} 
or decide that it does not have a solution in time 
$q^{3/5} (\log q)^{O(1)}$ as a quantum computer. 
\end{theorem}

\begin{proof}
Using Shor's algorithm~\cite{sh}, we can obtain the multiplicative orders 
$s_j$'s ($j=1,2,3$) in polynomial time. 
We may assume without loss of generality that $s_1 \geq s_2 \geq s_3$. 
As in the proof of Theorem \ref{th1}, we put $r$ as \eqref{r}. 
Further, we consider a polynomial time quantum subroutine 
$\mathcal{S} (x_2, x_3)$ which either finds and returns $x_1 \in X_1$ with 
\[ g_1^{x_1} = a_1^{-1} ( b-a_2 g_2^{x_2} -a_3 g_3^{x_3}) \]
or reports that no such $x_1$ exists for a given 
$(x_2, x_3) \in X_2 \times X_3 (r)$ by using Shor's discrete logarithm 
algorithm. 
\begin{enumerate}
\item The case $r \leq s_3$. 
Using Grover's search algorithm~\cite{g}, we search the subroutine 
$\mathcal{S} (x_2, x_3)$ for all $(x_2, x_3) \in X_2 \times X_3 (r)$ in time 
\[ (x_2 r)^{1/2} (\log q)^{O(1)} \ll q^{3/5} (\log q)^{O(1)}, \]
since $x_2 r \leq (s_1^2 s_2^2 r)^{2/5}$. 
\item The case $r > s_3$. 
Similarly, we search the $\mathcal{S} (x_2, x_3)$ for all 
$(x_2, x_3) \in X_2 \times X_3$ in time
\[ (x_2 x_3)^{1/2} (\log q)^{O(1)} \ll q^{3/5} (\log q)^{O(1)}, \]
since $x_2 x_3 \leq (s_1^2 s_2^2 s_3)^{2/5} < (s_1^2 s_2^2 r)^{2/5}$. 
\end{enumerate}
\end{proof}

In \cite{ds}, van Dam and Shparlinski mentioned when the multiplicative orders
are large, there is a more efficient quantum algorithm. Similarly, we can 
also consider a more efficient quantum algorithm. 

\begin{theorem} \label{th3}
If we assume 
\[ (s_1 s_2)^2 s_3 > q^3 \log q, \]
then we can either find a solution $(x_1, x_2, x_3) \in \bm{X}^3$ of the equation 
\eqref{th1-eq} or decide that it does not have a solution in time 
$q^{2} (s_1^2 s_2^2 s_3)^{-1/10} (\log q)^{O(1)}$ as a quantum computer, 
except for at most $q/\log q$ exceptional $b$'s. 
\end{theorem}

\begin{remark}
The upper bound of the running time of the algorithm of Theorem \ref{th3} is 
\[ O(q^{1/5} (\log q)^{O(1)}). \]
\end{remark}

\begin{proof}[\indent\sc Proof of Theorem \ref{th3}]
We may assume without loss of generality that $s_1 \geq s_2 \geq s_3$. 
We put 
\begin{equation}
r = \lfloor q^3 (s_1 s_2)^{-2} \log q \rfloor
\end{equation}
Then from the assumption of the theorem we see that $r \leq s_3$. Hence 
there are some solutions of \eqref{th1-eq} in $\bm{X}^3 (r)$ and 
we denote the number of the solutions of \eqref{th1-eq} by $M$. 
Note that $M \asymp (s_1 s_2 r)/q$. 

As in the case of \cite{ds}, we use the version of Grover's algorithm as 
described in \cite{bbht} that finds one out of $m$ matching items in a set 
of size $t$ by using only $O(\sqrt{t/m})$ queries. We search the subroutine 
$\mathcal{S} (x_2, x_3)$ for all $(x_2, x_3) \in X_2 \times X_3 (r)$. Then the 
time complexity is 
\[ \Bigl( \frac{s_2 r}{M} \Bigr)^{1/2} (\log q)^{O(1)} 
\leq q^{1/2} (s_1^2 s_2^2 s_3)^{-1/10} (\log q)^{O(1)}. \]
\end{proof}

%%%%%%%%%%%%%%%%%%%%%%%%%%%%%%%%%%%%%%%
\section{Concluding remarks}
%%%%%%%%%%%%%%%%%%%%%%%%%%%%%%%%%%%%%%%%%%%
At the end of this article, we compare the case of van Dam and Shparlinski with 
our case. See the following list. 
\begin{center}
\begin{tabular}{|c|c|c|c|} \hline
$\#$ of variables & Classical & Quantum & ratio (C/Q) \\ \hline
2 & 1 & 1/3 & 3 \\ \hline
3 & 3/2 & 3/5 & 5/2 \\ \hline
\end{tabular}
\end{center}
The case of two variables is that of van Dam and Shparlinski and 
the case of three variables is our case. 
We notice that the ratio decreases. Does the ratio decrease to $1$ 
when the dimension increase? We can apply the method used in \cite{ds} and 
this paper to the case of any variables. 
By roughly calculating, the ratio seems to converge $2$, when the number of 
the variables increases. 
It seems to come from the effect of Grover's algorithm.

\end{document}